\documentclass
[12pt,showpacs,preprintnumbers,nofootinbib,preprint,amsfonts,amssymb]{revtex4}%
\usepackage{amsmath}
\usepackage{graphicx}
\usepackage{bm}
\usepackage{amsfonts}
\usepackage{amssymb}
\usepackage{appendix}%
\begin{document}
\preprint{ }
\title{Triplet Dark Matter from leptogenesis}
\author{Jae Ho Heo}
\email{jaeheo1@gmail.com}
\author{C. S. Kim}
\email{cskim@yonsei.ac.kr}
\affiliation{Department of Physics and IPAP, Yonsei University, Seoul 120-479, Korea }

\begin{abstract}
A triplet dark matter candidate from thermal leptogenesis is considered with
building a model. The model is based on the standard two Higgs doublet model
and seesaw mechanism with Higgs triplets. The parameters (couplings and
masses) are adjusted for the observed small neutrino mass and the
leptogenesis. Dark matter particles can annihilate and decay in this model.
The time evolution of the dark matter number is governed by (co)annihilations
in the expanding universe, and its mass is constrained by the observed relic
density. The dark matter can decay into final states with three leptons (two
charged leptons and one neutrino). We investigate whether the decay in the
galaxy can account for cosmic ray anomalies in the positron and electron
spectrum. A noticeable point is that if the dark matter decays into each
lepton with different branching ratios, cosmic ray anomalies in AMS-02
measurements of the positron fraction and the Fermi LAT measurements of the
electrons-plus-positrons flux could be simultaneously accounted for from its
decay products. The leptogenesis within this model is studied in an appendix.

\end{abstract}

\maketitle

\section{Introduction}

Recent progress in cosmology and particle physics has eluded scientists for
more exact science. The Planck released data with relatively good precision,
and the standard model of particle physics has been tested by the discovery of
a Higgs-like boson with a mass around 126 GeV in both the ATLAS and CMS
experiments. Our current understanding of the universe is based on the
Friedman-Robertson-Walker (FRW) model and the standard model (SM) of particle
physics, called the standard cosmological model. Although we might understand
most of the observations in the standard cosmological model, dark matter (DM)
and baryon asymmetry in the universe (BAU) require new physics beyond the
standard (cosmological) model.

The DM and the BAU have quite appealing scenarios. Dark matter as a thermal
relic \cite{Rjs86} is well motivated in the hot big bang model. DM particles
would be in thermal equilibrium in the early universe and freeze out below its
mass scale in the expanding universe. The observed relic density \cite{Para13}
can naturally be explained by the annihilation cross section provided its mass
lies in the GeV-TeV range. The BAU may be explained if three conditions
proposed by Sakharov \cite{Ads67} are satisfied, namely baryon number
violation, C and CP violation and departure from thermal equilibrium in the
early universe. The most appealing candidate to explain the BAU must be
leptogenesis\footnote{The standard model tends to fail to realize the large
observed asymmetry because the only CP asymmetry is through the complex phase
in the Cabibbo-Kobayashi-Maskawa matrix and it is too small to explain the
observed baryon asymmetry. Furthermore a first order electroweak phase
transition is not plausible for Higgs mass with 126 GeV. Hence electroweak
baryogenesis is practically ruled out.} \cite{Mfu86}. The lepton asymmetry may
arise in the same dimension-five operator relevant to the neutrino mass. The
sphaleron processes convert a part of the lepton number to the baryon number,
and an excess of baryons can be explained.

In this paper, we utilize both properties with the additional particle content
in the standard model gauge group $SU(3)_{C}\times SU(2)_{L}\times U(1)_{Y}$.
A Majorana fermion triplet\footnote{The $\psi$ has three components $\left\{
\psi^{+},\psi^{0},\psi^{-}\right\}  $, and the neutral component is our DM
candidate. Since other components are in the same set, the $\psi$ is called
the triplet DM. In this paper, the symbol $\psi$ is also referred to as the DM
unless otherwise noted.} $\psi$ with a $SU(2)_{L}$ weak charge is considered
as a DM candidate with lifetime around $10^{26}\sec$. The seesaw mechanism
with a heavy triplet scalar (Higgs triplet) $\chi$ is employed to generate the
neutrino mass \cite{Rnm81} and the lepton asymmetry \cite{Ema98} by lepton
number violating interaction at the mass scale of $\chi$. We consider the
standard two Higgs doublet model (2HDM) as a low energy effective theory.

If our DM candidate is $Z_{2}$- odd, it will couple to $Z_{2}$-odd charged
leptons with the Higgs triplet. It can thus decay into three body final states
(two charged leptons and one neutral lepton) by a $\chi$ exchange. Since the
DM candidate has a weak charge, DM particles can also annihilate into SM
particles. The time evolution of DM number is governed by (co)annihilations in
the expanding universe, and its mass $(\sim2.7$ TeV$)$ is constrained by the
observed relic density. The decay process is negligible to the DM number
evolution and its lifetime is much longer than the age of the universe, 13.7
Gyr $(=4.3\times10^{17}\sec)$ within the $\Lambda$CDM concordance model
\cite{Par13}, the decay rate with lifetime around $10^{26}\sec$ is much larger
than annihilation rate at present. We examine whether cosmic ray anomalies in
the positron spectrum can be accounted for by DM decay. The predictions in
simple or single channels (democratic decay, $\mu^{+}\mu^{-}\nu$ dominant
decay, $\tau^{+}\tau^{-}\nu$ dominant decay) could fit each experimental
result of AMS-02 or Fermi LAT, but are unlikely to fit both experimental
results together. We calibrate our prediction by providing different branching
ratios into each channel. This method allows us to fit AMS-02 and Fermi LAT
measurements simultaneously. There are several models to accommodate the
decaying dark matter to account for the cosmic ray anomalies, dark matter in
grand unification models \cite{Ksb89,Dei89,Chc09,Aar09,Mil10,Car10}, sterile
neutrino dark matter \cite{Sdo94,Xsh98}, gravitino dark matter
\cite{Fta00,Wbu07,Slo07,Aib08,Kis08,Lco09,Koi09,Wbu09,Kyc10,Neb10,Kych10,Xia12,Gag12}%
, Goldstino dark matter \cite{Aib09,Hch12} and instanton-mediated dark matter
\cite{Chc10}.

The outline of this paper is as follows. In Sec. II we propose a model with a
triplet fermion. The seesaw mechanism and the standard 2HDM are employed. In
Sec. III we discuss the time evolution of the DM number density in the
expanding universe. In Sec. IV, the cosmic ray anomalies in the positron
spectrum are interpreted by DM decay. Finally, our conclusion is given in Sec.
V. In Appendix, baryon asymmetry via leptogenesis is studied within this model.

\section{The model description}

Our $Z_{2}$-odd DM candidate $\psi$ is completely stable in the SM. The only
interaction is an annihilation into SM particles through the operator
$\overline{\psi}{\not W  }\psi$. However, if we mind the seesaw mechanism with
at least a heavy Higgs triplet $\chi$ for tiny neutrino mass, the DM candidate
can have additional interactions in the standard 2HDM ($Z_{2}$ symmetric
2HDM). The standard 2HDM was built to avoid potentially large flavor changing
neutral currents (FCNCs) with $Z_{2}$ symmetry \cite{Slg77}, that is $d^{c}$,
$e^{c}$ and one Higgs doublet $\phi_{1}$ are $Z_{2}$-odd, and $u^{c}$ and the
other Higgs doublet $\phi_{2}$ are $Z_{2}$-even. Our $Z_{2}$-odd DM candidate
$\psi$ is thus allowed to couple to $Z_{2}$-odd charged leptons with the Higgs
triplet $\chi$. It can thus decay into three body final states by a $\chi$
exchange. The relevant potential which can describe interactions with new
particles is given by%

\begin{equation}
ig\overline{\psi}\not W\psi+y_{\psi}\mathrm{{Tr}}(\psi\chi^{\dag}%
)e^{c}+y_{\ell}\ell i\sigma_{2}\chi\ell+\mu_{1}\phi_{1}\chi i\sigma_{2}%
\phi_{1}+\mu_{2}\phi_{2}\chi i\sigma_{2}\phi_{2}+h.c.,
\end{equation}
where flavor indices are suppressed. The symbol $\ell$ stands for the
left-handed lepton doublet, and the components of Higgs doublets are $\left\{
\phi_{1,2}^{-},\phi_{1,2}^{0}\right\}  $ with gauge charge $\left(
1,2,-\frac{1}{2}\right)  $ in the gauge group $SU(3)_{C}\times SU(2)_{L}\times
U(1)_{Y}$. The fermion triplet $\left(  1,3,0\right)  $ and the Higgs triplet
$\left(  1,3,1\right)  $ were expressed in bilinear form,%

\[
\psi\equiv%
\begin{pmatrix}
\frac{1}{\sqrt{2}}\psi^{0} & \psi^{+}\\
\psi^{-} & -\frac{1}{\sqrt{2}}\psi^{0}%
\end{pmatrix}
,\chi\equiv%
\begin{pmatrix}
\frac{1}{\sqrt{2}}\chi^{+} & \chi^{++}\\
\chi^{0} & -\frac{1}{\sqrt{2}}\chi^{+}%
\end{pmatrix}
.
\]
The second term describes the lepton number violating interaction by one unit
$(\Delta L=1)$. The third term does the lepton number violating interaction by
two units $(\Delta L=2)$. The rest of the terms are scalar cubic potentials.

In the low energy effective theory, the heavy scalar triplet is decoupled. It
can be integrated out, and this handling gives rise to a sub-eV Majorana mass
of neutrinos as required by oscillation experiments. The tiny neutrino mass
can be generated by the combination of $\Delta L=2$ and Higgs cubic potentials,%

\begin{equation}
m_{\nu}\simeq y_{\ell}\frac{\left(  \mu_{1}v_{1}^{2}+\mu_{2}v_{2}^{2}\right)
}{2M_{\chi}^{2}},
\end{equation}
where $M_{\chi}$ is the mass of Higgs triplet, and $v_{1}/\sqrt{2}(v_{2}%
/\sqrt{2})$ is the vacuum expectation value of $\phi_{1}(\phi_{2})$. This form
is reduced to the usual standard form with $v^{2}=v_{1}^{2}+v_{2}^{2}%
\simeq246$ GeV for $\mu_{1}=\mu_{2}$. The strongest upper limit on the mass of
neutrinos comes from cosmology. The summed mass of the three neutrinos must be
less than 0.23 eV \cite{Para13} from the analysis of cosmological data such as
the cosmic microwave background radiation (CMB) and baryon acoustic
oscillations (BAO). On the other hand, there exists at least one neutrino mass
eigenstate with a mass of at least 0.04 eV \cite{Cam08} from atmospheric
neutrino oscillations. The mass scale of $\chi$ is of the order of
$10^{10}-10^{16}$ GeV, depending on the couplings $y_{\ell},$ $\mu_{1}$ and
$\mu_{2}$. The lepton asymmetry may arise in the lepton number violating
operators relevant to the neutrino mass $(\Delta L=2)$ and DM decay $(\Delta
L=1)$. The details of lepton asymmetry in this model are studied in Appendix.

\section{Dark matter annihilation and relic density}

In the expanding universe, the number density of DMs would decrease as long as
the temperature remains higher than the DM mass. When the temperature dropped
below the DM mass, the number density of DMs would drop exponentially
(Boltzmann suppression). If equilibrium was maintained until today, there
would be very few DMs left, but the DM number density would freeze out at some
point and a substantial number of DMs would be left today. Detailed evolution
of the Boltzmann equation is necessary for an accurate prediction. In our
model, the DM can decay and annihilate. The time evolution Boltzmann equation
of DM number density is given by%

\begin{equation}
Y^{\prime}(x)=-\frac{\Gamma}{xH}\left(  Y-Y_{\text{eq}}\right)  -\frac
{s\left\langle \sigma_{\text{eff}}v\right\rangle }{xH}\left(  Y^{2}%
-Y_{\text{eq}}^{2}\right)
\end{equation}
where $x=M/T$ is the inverse temperature with DM mass $M$, $Y(Y_{\text{eq}})$
is the (equilibrium) number density in units of entropy density $s,$ $H$ is
the Hubble parameter, $\Gamma$ is the DM decay rate (width) and $\left\langle
\sigma_{\text{eff}}v\right\rangle $ is the effective annihilation cross
section. We defined the $^{\prime}$ notation as $^{^{\prime}}\equiv\left(
1-\frac{x}{4}\frac{d\ln g_{\ast}(x)}{dx}\right)  ^{-1}\frac{d}{dx}$ with the
effective relativistic degrees of freedom $g_{\ast}(x)$ which is constant in
the adiabatic expansion universe. If we consider only the decay part of the
Boltzmann equation after freeze-out, DM particles are approximately decreasing
with the rate $1-\exp\left(  -\Gamma/2H(x)\right)  $ in number. Otherwise,
they are decreasing with the rate $s\left\langle \sigma_{\text{eff}%
}v\right\rangle /H$ in number for annihilation. The decreasing rate by
annihilation $\left\langle \sigma_{\text{eff}}v\right\rangle \sim10^{-26}$
cm$^{3}\sec^{-1}$ is much larger than the one by decay $\Gamma\sim10^{-26}%
\sec^{-1}$. For example, the decreasing rate will be $10^{-11}$ by decay and
$10^{-6}$ by annihilation in the present day universe $H_{0}\sim10^{-16}%
\sec^{-1},s_{0}\sim3000$ cm$^{-3}$. The difference must be much larger at
freeze-out. The annihilation dominantly contributes to the time evolution of
the DM number density. We thus neglect the contribution of DM decay to the
time evolution Boltzmann equation. These small decreasings must be negligible
to other astrophysical and cosmological observations as well.

The triplet DM has three components $\left\{  \psi^{+},\psi^{0},\psi
^{-}\right\}  $, and each component must have the similar thermal history and
be nearly degenerate. We need include coannihilation effects in the
calculation of the relic density. The coannihilation effects can be described
in the effective cross section \cite{Kgr91} with the following form%

\begin{equation}
\sigma_{\text{eff}}=\sum_{i,j}\sigma_{ij}\frac{g_{i}g_{j}}{g_{\text{eff}}^{2}%
}\left(  1+\Delta_{i}\right)  ^{3/2}\left(  1+\Delta_{j}\right)
^{3/2}e^{-x\left(  \Delta_{i}+\Delta_{j}\right)  },
\end{equation}
where $\Delta_{i}=\left(  M_{i}-M\right)  /M$ $\left(  i=+,0,-\text{ and
}M=M_{0}\right)  ,$ $g_{\text{eff}}=%
{\textstyle\sum}
g_{i}\left(  1+\Delta_{i}\right)  ^{3/2}\exp\left(  -x\Delta_{i}\right)  $
with $g_{i}$ internal degrees of freedom of DM components and $\sigma_{ij}$ is
the cross section between $i$ and $j$. Four processes are related to the
calculation of the effective cross section, $\psi^{0}\psi^{0},\psi^{+}\psi
^{-},\psi^{\pm}\psi^{0},\psi^{\pm}\psi^{\pm}$ annihilations. The mass
difference between our DM components are $160-170$ MeV \cite{Jlf99}. For such
small mass difference, $\Delta_{i,j}$ are negligible. The effective cross
section $\sigma_{\text{eff}}$ becomes the average of all relevant cross
sections in this case, and we get the effective annihilation cross section
$\left\langle \sigma_{\text{eff}}v\right\rangle \simeq3\pi\alpha_{g}^{2}%
/M^{2}$ where $\alpha_{g}=g^{2}/4\pi$ is the weak fine structure constant.
From the Boltzmann equation (3) with the relation $Y=Y_{+}+Y_{0}+Y_{-}$, the
DM relic density $(\Omega_{\text{DM}}h^{2}\simeq0.12)$ can be, according to
the study of wino DM in \cite{Jun08} and minimal DM in \cite{Mci09} for
annihilations through the operator $\overline{\psi}{\not W  }\psi$, explained
with DM mass around 2.7 TeV.

\section{Dark matter decay and cosmic ray signals}

The DM decay and annihilation into SM particles in the universe would
contribute to the observed cosmic rays. The decay rate $(\Gamma\sim
10^{-26}\sec^{-1})$ is larger than the annihilation rate $(n_{\text{DM}%
}\left\langle \sigma v\right\rangle \sim10^{-31}\sec^{-1})$ at present. The
contribution of DM decay to the cosmic rays are considered. The DM can decay
into three body final states through the lepton number violating interaction,
and we get interested in the decay mode $\psi\longrightarrow e_{i}^{+}%
e_{j}^{-}\nu_{j}\left(  e_{i}^{-}e_{j}^{+}\overline{\nu}_{j}\right)  $ where
$i,j$ are flavor indices as depicted in Fig. 1. The decay rate results in%

\begin{equation}
\Gamma=%
{\textstyle\sum\limits_{i,j}}
\frac{1}{64\pi^{3}M}\int_{0}^{\frac{1}{2}M}dE_{1}\int_{\frac{1}{2}M-E_{1}%
}^{\frac{1}{2}M}dE_{2}\left\langle \left\vert \mathcal{M}\right\vert
^{2}\right\rangle =%
{\textstyle\sum\limits_{i,j}}
\frac{y_{\psi_{i}}^{2}y_{\ell j}^{2}}{6144\pi^{3}}\frac{M^{5}}{M_{\chi}^{4}},
\end{equation}
where $\mathcal{M}$ is the scattering amplitude for this decay process and the
angle bracket means averaging over initial spins and summing over final spins.
All the final states are assumed to be massless. Notice that the maximum
energy which a produced particle can have is $M/2$. The DM lifetime is%

\begin{equation}
\tau_{\text{DM}}=\Gamma^{-1}\simeq10^{26}\sec\left(  \frac{2700\text{ GeV}}%
{M}\right)  ^{5}\left(  \frac{M_{\chi}}{10^{15}\text{ GeV}}\right)  ^{4}%
\frac{(0.3)^{2}(0.3)^{2}}{%
{\textstyle\sum\limits_{i,j}}
(y_{\psi i})^{2}(y_{\ell j})^{2}}.
\end{equation}
As far as Yukawa couplings are not seriously fine-tuned, the lifetime is of
the order of $10^{26}\sec$ for Higgs triplet mass around $10^{15}$ GeV.%

\begin{figure}
[ptb]
\begin{center}
\includegraphics[
trim=0.000000in 0.000000in -0.014436in 0.000000in,
height=2.9413cm,
width=5.5004cm
]%
{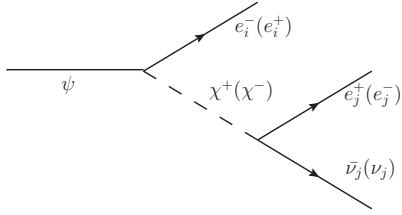}%
\caption{The Feynman diagram of dark matter decay.}%
\end{center}
\end{figure}

Recently, the cosmic ray anomalies more clearly appeared in the positron
spectrum. The AMS-02 \cite{Mag13} has observed a steep rise of the positron
fraction over the theoretical expectation up to 350 GeV in kinetic energy, and
the PAMELA \cite{Oad13} made new measurements with a steep rise that extend
the previous measurements \cite{Oad09} up to 300 GeV. The AMS-02 data show
much higher precision and wider energy extension. Their results must be
consistent in their systematic errors, however the spectrum of AMS-02 tends to
be softer. Both results must require additional sources of their origin in the
galaxy.\ An excess over the theoretical prediction also appeared in
electrons-plus-positrons measurements at the Fermi LAT \cite{Mac10} up to
$\sim1-2$ TeV in kinetic energy, combined with HESS results \cite{Fah08,Fah09}.%

\begin{figure}
[ptb]
\begin{center}
\includegraphics[
trim=0.000000in 0.000000in -0.003409in 0.000000in,
height=17.3138cm,
width=16.0595cm
]%
{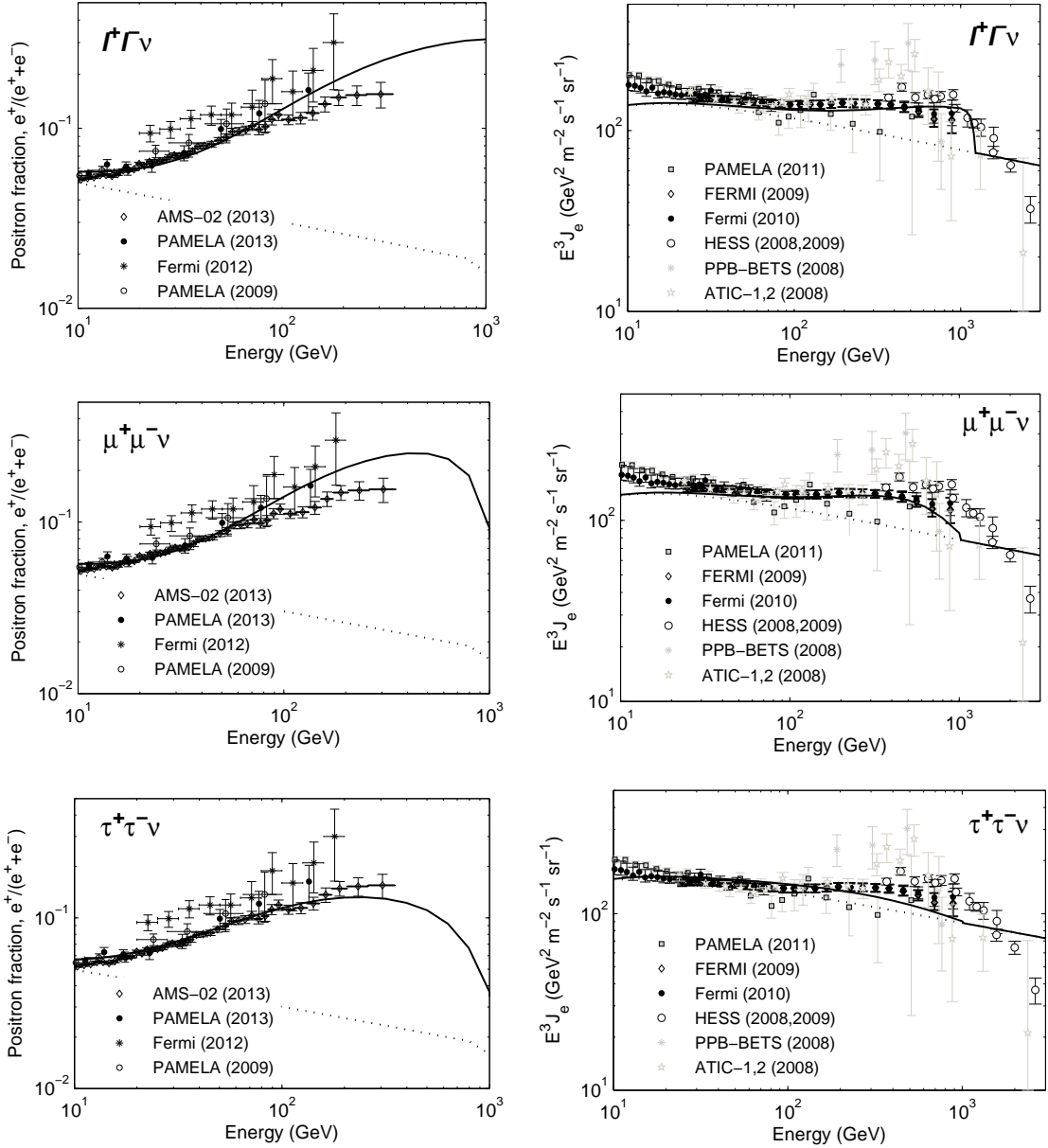}%
\caption{Predicted cosmic ray signals in $l^{+}l^{-}\nu,\mu^{+}\mu^{-}\nu
,\tau^{+}\tau^{-}\nu$ decay channels with DM mass 2.5 TeV$.$Left panels:
Positron fraction with experimental data, AMS-02 \cite{Mag13}, PAMELA
\cite{Oad13,Oad09}, Fermi LAT \cite{Mac12}. Right panels:
Positrons-plus-electrons flux with experimental data, PAMELA (electron only)
\cite{Oad11}, Fermi LAT \cite{Mac10}, HESS \cite{Fah08,Fah09}, PPB-BETS
\cite{Sto08}, and ATIC \cite{Jch08}. The bold dotted line shows the
astrophysical background. Solar modulation is taken into account by using the
force field approximation with the Fisk potential 600 MV.}%
\end{center}
\end{figure}

In Fig. 2, we show the predicted positron fraction and the
electrons-plus-positrons flux by DM decay with mass 2.5 TeV. The predictions
are made for the democratic decay with a universal coupling $(l^{+}l^{-}\nu)$,
muon dominant decay $(\mu^{+}\mu^{-}\nu)$ and tauon dominant decay $(\tau
^{+}\tau^{-}\nu)$. The primary electron flux of the astrophysical background
is from PAMELA electron flux fit \cite{Oad11} with the spectral index $-3.18$
(injection index:$-2.66$) above the energy region influenced by the solar wind
($\geq30$ GeV). The secondary positron flux of the background is from the
GALPROP conventional model \cite{Aws04} in the analytic form \cite{Aib10}. The
density profile of the Milky Way halo is adopted to be the Navarro-Frenk-White
(NFW) distribution \cite{Jfn96} and the MED propagation model \cite{Dma01} is
selected for galactic cosmic ray transport. The similar plots exist in Ref.
\cite{Aib10} with various DM masses, and Ref. \cite{Kko13} for tauon dominant
decay $(\tau^{+}\tau^{-}\nu)$ with DM mass of 3 TeV. The predictions must be
very similar. Our predictions of $l^{+}l^{-}\nu$ with lifetime $5.6\times
10^{26}\sec$ and $\mu^{+}\mu^{-}\nu$ with lifetime $1.7\times10^{26}\sec$ are
likely to fit both PAMELA results of the positron excess and Fermi LAT
measurements of electrons-plus-positrons flux simultaneously, but they are in
tension with AMS-02 energy spectrum above 100 GeV. Otherwise, the prediction
of $\tau^{+}\tau^{-}\nu$ \ with lifetime $1.2\times10^{26}\sec$ is likely to
fit the AMS-02 result, but it cannot explain the Fermi LAT measurements. It
has already been noticed a difficulty on fitting the AMS-02 and Fermi LAT
results together, and there are studies on how to relax the tension
\cite{Lfe13}.%

\begin{figure}
[ptb]
\begin{center}
\includegraphics[
trim=0.000000in 0.000000in -0.827263in 0.000000in,
height=8.7931cm,
width=16.0595cm
]%
{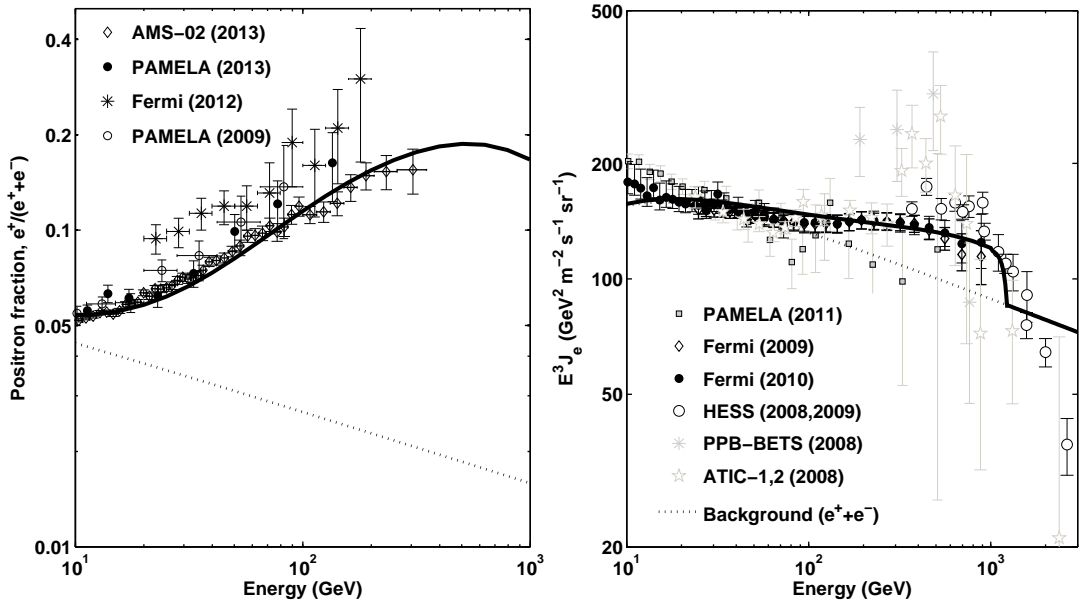}%
\caption{Same as Fig. 2, but dark matter decay with branching ratios,
$B_{e}=6\%,B_{\mu}=6\%\ $and $B_{\tau}=88\%$, and life time $2.0\times
10^{26}\sec.$}%
\end{center}
\end{figure}

In this work we calibrate predictions by providing different branching ratios
in each channel. In most studies, a simple or single channel has been adopted
to fit AMS-02 and Fermi LAT results simultaneously such that we did in Fig. 2.
Our prediction in $l^{+}l^{-}\nu$ and $\mu^{+}\mu^{-}\nu$ channels is much
harder than the AMS-02 result above 100 GeV, otherwise the prediction in the
$\tau^{+}\tau^{-}\nu$ channel is softer. In electrons-plus-positrons spectrum,
our prediction in the $l^{+}l^{-}\nu$ channel shows a sharp feature near the
maximum energy, otherwise the prediction in the $\tau^{+}\tau^{-}\nu$ channel
is very soft near the energy. If our DM can decay into each lepton with
different branching ratios, it is possible that we make an appropriate fit of
AMS-02 and Fermi LAT measurements together. We show an appropriate fit in Fig.
3 with branching ratios, $B_{e}=6\%,B_{\mu}=6\%\ $and $B_{\tau}=88\%$ and the
lifetime $2.0\times10^{26}\sec$. The predictions with different branching
ratios are likely to fit AMS-02 and Fermi LAT measurements together. Other
divisions of the branching ratio might provide better fits\footnote{Flavor
mixing channels such as $e^{+}\mu^{-}\nu_{\mu},$ $\mu^{+}\tau^{-}\nu_{\tau}$
and \ $\tau^{+}e^{-}\nu_{e}$ are also possible in our model. Predictions in
each flavor mixing channel are, according to Ref. \cite{Mib13}, unlikely to
fit AMS-02 and Fermi LAT measurements together. We might consider a
calibration with the flavor mixing channels. However, the spectra are
dominantly determined by the spallation of incident particles in the order
$e^{-}(e^{+}),\mu^{-}(\mu^{+})$ and $\tau^{-}(\tau^{+})$, and so there would
be no big difference from predictions in flavor conserving channels. For
example, the spectra in $\mu^{+}\mu^{-}\nu_{\mu}$ and $\mu^{+}\tau^{-}%
\nu_{\tau}$ channels are determined by the spallation of $\mu^{+}$. The flavor
mixing channels are just involved in the detailed spectral shape.}.

\section{Conclusions}

We proposed a triplet dark matter model based on the standard two Higgs
doublet model and seesaw mechanism with Higgs triplets. The lepton asymmetry
arises through the operators relevant to the neutrino mass $(\Delta L=2)$ and
dark matter decay $(\Delta L=1)$. Our dark matter candidate can annihilate and
decay into SM particles. The time evolution of the dark matter number is
governed by (co)annihilations in the expanding universe, and its mass is
constrained by the observed relic density. The dark matter is no longer
stable, and can slowly decay into three body final states (two charged leptons
and one neutrino). The decay products would contribute to the observed comic
rays, and they are able to explain cosmic ray anomalies in the positron
spectrum observed at AMS-02, PAMELA and Fermi LAT. A noticeable point is that
if dark matter particles decay into each lepton with different branching
ratios, cosmic ray anomalies in AMS-02 results of the positron fraction and
the Fermi LAT measurements of the electrons-plus-positrons flux could be
simultaneously accounted for from its decay products.

\begin{acknowledgments}
This work was supported by the National Research Foundation of Korea (NRF)
grant funded by Korea government of the Ministry of Education, Science and
Technology (MEST) (No. 2011-0017430) and (No. 2011-0020333).
\end{acknowledgments}

\appendix*

\section{Leptogenesis}

A lepton asymmetry can be generated in the decay of Higgs triplet $\chi$ if
the number of Higgs triplets is two or more. We re-express the Lagrangian of
Eq. (1) in the form%

\begin{equation}
y_{\psi k}\mathrm{{Tr}}(\psi\chi_{k}^{\dag})e^{c}+y_{\ell_{k}}\ell i\sigma
_{2}\chi_{k}\ell+\mu_{1k}\phi_{1}\chi_{k}i\sigma_{2}\phi_{1}+\mu_{2k}\phi
_{2}\chi_{k}i\sigma_{2}\phi_{2}+h.c.,
\end{equation}
where $k=1,2$ is a species index of $\chi$. If there is only one $\chi$, the
relative phase among couplings $y_{\psi k}{,}y_{\ell_{k}},\mu_{1k}$ and
$\mu_{2k}$ can be chosen real. There would be no CP-violating interaction.
With two $\chi^{\prime}$s, two relative phases must remain among the
couplings. A lepton asymmetry is dynamically generated by the interference
between the tree and one-loop level decay amplitudes, as shown in Fig. 4.
There is no one loop vertex correction. In general, the mass of $\chi^{\prime
}$s is different. The heavy particle $\chi_{2}$ would decay at higher
temperature (earlier time), and the lepton asymmetry by decay of $\chi_{2}$
will be washed out by the lepton number violating interaction of the light
particle $\chi_{1}$. Hence we only consider the lepton asymmetry by decay of
the light one $\chi_{1}$.%

\begin{figure}
[ptb]
\begin{center}
\includegraphics[
trim=0.000000in 0.000000in -1.110907in 0.000000in,
height=2.4492cm,
width=11.6838cm
]%
{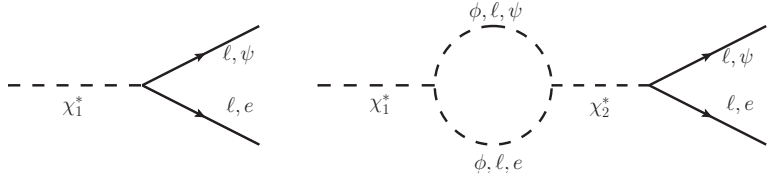}%
\caption{The decay of $\chi_{1}^{\ast}\longrightarrow\ell\ell,\psi e$ at tree
level and in one-loop order. A lepton asymmetry is generated by their
interference.}%
\end{center}
\end{figure}

The lepton asymmetry per decay (net lepton number) is defined by the
difference between the decay of $\chi_{1},\chi_{1}^{\ast}$ particles,%

\begin{align}
\delta_{l}  &  =2\left[  B\left(  \chi_{1}^{\ast}\rightarrow ll\right)
-B\left(  \chi_{1}\rightarrow l^{c}l^{c}\right)  \right]  ,\\
\delta_{\psi}  &  =B\left(  \chi_{1}^{\ast}\rightarrow l\psi\right)  -B\left(
\chi_{1}\rightarrow l^{c}\psi\right)  .
\end{align}
Since our DM couples to charged leptons with $\Delta L=1$, we have an
additional contribution. The $\delta_{l}$ is different from $\delta_{\psi}$ by
the factor 2, because two leptons are produced per decay. The total lepton
asymmetry will be $\delta_{L}=\delta_{l}+\delta_{\psi}$.

If we employ the procedure of Ref. \cite{Ema98} for detailed calculations of
the lepton asymmetry, the lepton asymmetry per decay results in%

\begin{align}
\delta_{l}  &  =\frac{1}{8\pi^{2}}\frac{\operatorname{Im}\left[  \left(
\mu_{11}\mu_{12}^{\ast}+\mu_{21}\mu_{22}^{\ast}+M_{1}M_{2}y_{\psi1}y_{\psi
2}^{\ast}\right)  y_{\ell1}y_{\ell2}^{\ast}\right]  }{M_{2}^{2}-M_{1}^{2}%
}\left[  \frac{M_{1}}{\Gamma_{1}}\right]  ,\\
\delta_{\psi}  &  =\frac{1}{16\pi^{2}}\frac{\operatorname{Im}\left[  \left(
\mu_{11}\mu_{12}^{\ast}+\mu_{21}\mu_{22}^{\ast}+M_{1}M_{2}y_{\ell1}y_{\ell
2}^{\ast}\right)  y_{\psi1}y_{\psi2}^{\ast}\right]  }{M_{2}^{2}-M_{1}^{2}%
}\left[  \frac{M_{1}}{\Gamma_{1}}\right]  ,
\end{align}
where $M_{1,2}$ are masses of $\chi_{1,2}$. All the final states are assumed
to be massless. The notations we used are $y_{\ell1}y_{\ell2}^{\ast}%
=\sum\limits_{i,j=e,\mu,\tau}y_{\ell1ij}y_{\ell2ij}^{\ast}$ and $y_{\psi
1}y_{\psi2}^{\ast}=\sum\limits_{i=e,\mu,\tau}y_{\psi1i}y_{\psi2i}^{\ast}$. The
triplet decay width at tree level is%

\begin{equation}
\Gamma_{1}=\frac{M_{1}}{8\pi^{2}}\left(  y_{\ell}y_{\ell}^{\ast}+y_{\psi
}y_{\psi}^{\ast}+\frac{\mu_{1}\mu_{1}^{\ast}+\mu_{2}\mu_{2}^{\ast}}{M_{1}^{2}%
}\right)
\end{equation}
in the notations, $y_{\ell}y_{\ell}^{\ast}=\sum\limits_{i,j=e,\mu,\tau}%
y_{\ell1ij}y_{\ell1ij}^{\ast},$ $y_{\psi}y_{\psi}^{\ast}=\sum\limits_{i=e,\mu
,\tau}y_{\psi1i}y_{\psi1i}^{\ast}$ and $\mu_{1}\mu_{1}^{\ast}+\mu_{2}\mu
_{2}^{\ast}=\mu_{11}\mu_{11}^{\ast}+\mu_{21}\mu_{21}^{\ast}$.

We consider the lepton asymmetry for Higgs triplets with mass around $10^{15}$
GeV in this work. If the decay is slower than the expansion rate of the
universe at temperature $T\sim M_{1}\sim10^{15}$ GeV, the $\chi,\chi^{\ast}$
bosons do not decrease in number till $t\sim\Gamma_{1}^{-1}$, and the number
density of $\chi,\chi^{\ast}$ bosons is $n_{\chi}=n_{\chi^{\ast}}\sim
n_{\gamma}$ where $n_{\gamma}$ is the number density of photons. Since each
decay produces a lepton number $\delta_{L}$, the lepton number density results
in $n_{L}\sim\delta_{L}n_{\chi}\sim\delta_{L}n_{\gamma}$. The produced lepton
asymmetry will be $Y_{L}=n_{L}/s\sim\delta_{L}/g_{\ast}$ with the entropy
density $s\sim g_{\ast}n_{\gamma}$. At a temperature above electroweak phase
transition, a part of the lepton asymmetry gets converted to the baryon
asymmetry via the $SU(2)_{L}$ sphaleron processes \cite{Jah90}, $Y_{B}%
=S_{B}Y_{L}\sim S_{B}\delta_{L}/g_{\ast}$. With $g_{\ast}\sim100$ and
$S_{B}\sim0.5$, the baryon asymmetry $Y_{B}\sim10^{-10}$ could be accounted
for by $\delta_{L}\sim10^{-8}-10^{-7}$. This small value of $\delta_{L}$ is
easily acquired from Eqs. (A.4,5). Although the mass of these triplets is
around $10^{15}$ GeV, there is still a possibility that the decay is faster
than the expansion rate of the universe. In this case, the lepton asymmetry
will be approximately suppressed by factor $1/K\left(  \ln K\right)  ^{0.6}$
\cite{Ewk94} where $K=\Gamma_{1}/H$. Near $K=1$, this suppression would be
easily restored by a slight enhancement of $\delta_{L}$. Otherwise, the
detailed time evolution Boltzmann equations are required in order to predict
the exact lepton asymmetry for $K\ll1$.

\end{document}